\documentclass[conference]{IEEEtran}
\IEEEoverridecommandlockouts

\usepackage{amsmath}
\usepackage{amssymb}
\usepackage{graphicx}
\usepackage{booktabs}
\usepackage{cite}
\usepackage{siunitx}
\usepackage{float}
\usepackage{multirow}
\usepackage[table]{xcolor}

\title{Attack-Dependent Robustness of Neural Audio Codecs for Adversarial ASR}

\author{
\IEEEauthorblockN{
Jordan Prescott, Thanathai Lertpetchpun, and Shrikanth Narayanan
}
\IEEEauthorblockA{
Signal Analysis and Interpretation Laboratory\\
University of Southern California, USA
}}

\begin{document}

\maketitle

\begin{abstract}
Neural audio codecs impose a discrete bottleneck through residual vector quantization (RVQ), making them a useful class of inference-time transformations for reducing adversarial perturbations before ASR inference. We study how codec quantization depth affects defended ASR under non-adaptive, standard adaptive, and quantization-aware adaptive untargeted $\ell_\infty$ attacks. Under non-adaptive attacks, intermediate RVQ depths yield the lowest word error rates and outperform traditional compression at comparable bitrates. However, this apparent optimum is not stable under adaptive evaluation. The standard identity-gradient adaptive baseline (BPDA+EOT) can overestimate robustness, while an implementation of an RVQ-relaxed adaptive attack (SoftVQ-PGD) substantially changes the observed depth trend and largely removes the intermediate-depth advantage. Overall, neural codecs can improve defended ASR under specific threat models. However, the relationship between robustness and RVQ depth depends on the attack used for evaluation, rather than on the codec architecture alone.
\end{abstract}


\section{Introduction}
Automatic speech recognition (ASR) systems are vulnerable to adversarial attacks, where small, carefully crafted perturbations induce incorrect or malicious transcriptions while preserving the original linguistic content for human listeners~\cite{carlini2018audio,qin2019imperceptible}. Inference-time preprocessing defenses are attractive because they can be applied before ASR without retraining or modifying the recognition model~\cite{lan2022adversarial,athalye2018obfuscated,andronic2020mp3,joshi2021study}. However, such defenses are difficult to evaluate. Reductions in attack success may reflect genuine removal of adversarial structure, or they may arise because the transformation obstructs gradient-based optimization under adaptive attack~\cite{athalye2018obfuscated}. This distinction is practically important: a codec configuration selected under non-adaptive or weak adaptive evaluation may not remain robust against an attacker that accounts for the transformation.

Recent work has explored learned audio transformations for downstream robustness. Chen et al.~\cite{chen2024neural} studied neural codecs for adversarial sample detection in speaker verification, while Ozer et al.~\cite{ozer2026selfvoiceconversionattack} showed that voice conversion models can disrupt neural audio watermarking while preserving linguistic content. These findings motivate examining whether robustness conclusions for learned audio transformations depend on the evaluation methodology.

Neural audio codecs (NAC) are particularly useful for studying this question because they introduce a discrete bottleneck with controllable resolution. Modern neural codecs discretize latent representations via residual vector quantization (RVQ)~\cite{defossez2022high,kumar2023high,defossez2024moshi}, where each codebook quantizes the residual left by the previous stage. Shallow depths impose stronger compression and may remove adversarial detail at the cost of speech fidelity, while deeper depths better preserve speech but may also preserve attack-relevant structure. RVQ depth therefore provides a controlled probe for analyzing whether codec robustness trends are stable properties of the transformation or artifacts of the attack used to evaluate it.

Because RVQ depth changes both compression strength and reconstruction
fidelity, we use it as an analysis axis rather than proposing a fixed
neural codec defense. We ask whether robustness conclusions drawn under non-adaptive evaluation remain stable when attacks account for the defended codec-ASR pipeline. To do so, we compare non-adaptive Projected Gradient Descent (PGD)~\cite{madry2018towards}, which optimizes against the ASR model alone, with Backward Pass Differentiable Approximation with Expectation over Transformation (BPDA+EOT)~\cite{athalye2018obfuscated}, which uses an identity-gradient approximation through the codec, and SoftVQ-PGD, which uses a differentiable relaxation of RVQ assignment while retaining hard quantization at evaluation~\cite{agustsson2017soft,jang2017categorical,bengio2013estimating}. Our results show that under untargeted $\ell_\infty$ evaluation, robustness trends over RVQ depth depend on the attack used to evaluate the defended codec-ASR pipeline, rather than being intrinsic to codec architecture alone.

Within this threat model, our main findings are:
\begin{enumerate}

\item RVQ depth exhibits an apparent U-shaped word error rate (WER) trade-off under non-adaptive PGD. Intermediate depths outperform traditional compression at comparable bitrates, while shallow and deep configurations are less effective in this threat model.

\item BPDA+EOT can give optimistic adaptive robustness estimates for
RVQ-based transformations because its identity-gradient surrogate shows
weak optimization through the discrete bottleneck.

\item A differentiable relaxation of RVQ quantization, which we call SoftVQ-PGD, provides a stronger quantization-aware adaptive attack by optimizing through codebook-distance geometry rather than an identity approximation.

\item SoftVQ-PGD substantially changes the RVQ-depth trend observed under PGD, largely eliminating the intermediate-depth optimum. Thus, the apparent optimal codec depth is dependent on the attack used to evaluate the defended codec-ASR pipeline.

\end{enumerate}

\section{Background}

\subsection{Neural Audio Codecs}
Neural audio codecs implement learned encoder--decoder architectures that compress waveforms through a discrete latent bottleneck. An encoder $E(\cdot)$ maps an input waveform $x$ to a latent representation that is quantized and reconstructed by a decoder $G(\cdot)$, enabling low-bitrate compression while preserving perceptual quality.

Many neural codecs use RVQ to discretize latent features~\cite{defossez2022high, kumar2023high, defossez2024moshi}. Let $r_1 = E(x)$ denote the encoder output. RVQ applies a sequence of $N$ codebooks, each of which quantizes the residual $r_k$ left by the previous stage:

\begin{equation}
q_k = Q_k(r_k), \qquad r_{k+1} = r_k - q_k,
\end{equation}
yielding the quantized representation $z_q = \sum_{k=1}^{N} q_k$ and reconstruction $\hat{x}=G(z_q)$. Each hard quantizer selects the nearest codeword from a learned codebook $\mathcal{C}_k$:
\begin{equation}
Q_k(r_k) = \arg\min_{e \in \mathcal{C}_k} \|r_k - e\|_2 .
\end{equation}
The number of codebooks $N$ controls reconstruction fidelity: larger $N$ reduces quantization error while smaller $N$ enforces coarser quantization that suppresses fine-grained signal variation. Because hard nearest-neighbor assignment is piecewise constant, the quantization step has zero gradient almost everywhere, making standard backpropagation through the RVQ bottleneck ineffective without a surrogate.

\subsection{Adversarial Attacks}
\label{sec:background_adv_attacks}

Adversarial examples are inputs modified with small, carefully crafted perturbations that cause a model to produce incorrect outputs. Given an input signal $x$ and a model $f(\cdot)$, untargeted adversarial attacks seek a perturbation $\delta$ within a bounded norm constraint that maximizes transcription error:
\begin{equation}
\max_{\delta \in \mathcal{B}_p(\epsilon)} \mathcal{L}(f(x+\delta), y),
\end{equation}
where $\mathcal{B}_p(\epsilon)$ denotes the $\ell_p$ ball of radius $\epsilon$, $\mathcal{L}(\cdot)$ is the ASR loss function, and $y$ is the ground-truth transcription. In ASR, untargeted attacks increase WER by maximizing decoding loss, while targeted attacks minimize the loss toward a predefined target phrase.

When a preprocessing defense $C(\cdot)$ is applied at inference, the attacker's knowledge of and access to that defense determines the threat model. In the \emph{non-adaptive} setting, the perturbation is optimized against $f(\cdot)$ alone and the defense is applied only at evaluation. PGD~\cite{madry2018towards} is the standard non-adaptive baseline: it iteratively updates $\delta$ using the ASR loss gradient and projects onto the $\ell_\infty$ ball, with no knowledge of the defense being used.

In the \emph{adaptive} setting, the attacker optimizes $\delta$ through the full defended pipeline $f(C(\cdot))$. Because RVQ quantization is non-differentiable, this requires a gradient surrogate, and the strength of the resulting attack depends directly on the quality of that surrogate. BPDA+EOT~\cite{athalye2018obfuscated} is a common approach that replaces $\partial C / \partial x$ with the identity during backpropagation---an instance of the straight-through estimator~\cite{bengio2013estimating}---EOT reduces gradient variance by averaging the loss over Gaussian-jittered inputs:
\begin{equation}
\max_{\delta \in \mathcal{B}_p(\epsilon)}
\mathbb{E}_{\eta \sim \mathcal{N}(0,\sigma^2)}
\left[
\mathcal{L}\big(f(C(x + \delta + \eta)), y \big)
\right],
\end{equation}
where $C(\cdot)$ denotes a transformation, like RVQ-based codecs.

The use of the identity surrogate is computationally convenient but ignores the structure of vector quantization. It does not differentiate through codebook selection, which can yield weak or misleading optimization signals when the defense relies on a discrete bottleneck~\cite{athalye2018obfuscated}.

For RVQ-based defenses, stronger gradient-based adaptive evaluation can use surrogates that approximate the quantization step itself. A natural approach replaces the hard nearest neighbor assignment with a temperature-scaled softmax over codebook distances, yielding differentiable weighted reconstructions while retaining hard quantization at evaluation~\cite{agustsson2017soft, bengio2013estimating}. Related quantization-aware attacks have been studied for product-quantized image retrieval systems~\cite{feng2020adversarialattackdeepproduct}. We adapt this idea to RVQ-based neural audio codec transformations and use the resulting differentiable relaxation as an adaptive attack, which we refer to as SoftVQ-PGD.

\subsection{Existing Defense Mechanisms}
Defenses against adversarial attacks in speech recognition are often grouped into three categories: adversarial training, detection-based methods, and input preprocessing~\cite{zhang2022adversarial,zhao2024adversarial}. Adversarial training incorporates adversarial examples during model optimization to improve robustness~\cite{madry2018towards,mehlman2023mel,9414843,JATI2021101199}, but requires increased computational cost and typically improves robustness only within the perturbation regimes seen during training. Detection-based approaches aim to identify manipulated inputs without modifying the ASR model~\cite{jayashankar20_interspeech,lan2022adversarial}. However, they do not remove perturbations and may be bypassed by adaptive attacks. Preprocessing defenses apply signal transformations prior to inference, including compression, filtering, learned enhancement, or other waveform manipulations~\cite{chen2024neural,andronic2020mp3,hussain2021waveguard,chen2023towards,raber2025keep,joshi2021study,zelasko2021adversarial, Yang_2020}. These methods are attractive because they operate at inference time and require no retraining, but can fail when attackers explicitly account for them~\cite{athalye2018obfuscated}.

Our work belongs to the preprocessing category. Rather than evaluating a single fixed transformation, we use RVQ depth as an analysis axis across multiple neural codec families and study how attack formulation changes the apparent robustness of the defended codec-ASR pipeline.

\section{Methodology}

\subsection{Codec Transformation and RVQ Depth}
We model the defended ASR pipeline as $f(C_N(x))$, where $f$ is the ASR model and $C_N$ is a neural codec reconstruction using the first $N$ RVQ codebooks. Varying $N$ changes the effective resolution of the discrete bottleneck: smaller $N$ enforces coarser reconstruction, while larger $N$ preserves finer signal detail. We use $N$ as the primary controlled variable for measuring how codec capacity affects adversarial degradation and adaptive attack behavior.

\subsection{Threat Model and Attacks}
\begin{figure}[t]
    \centering
    \includegraphics[width=0.8\linewidth]{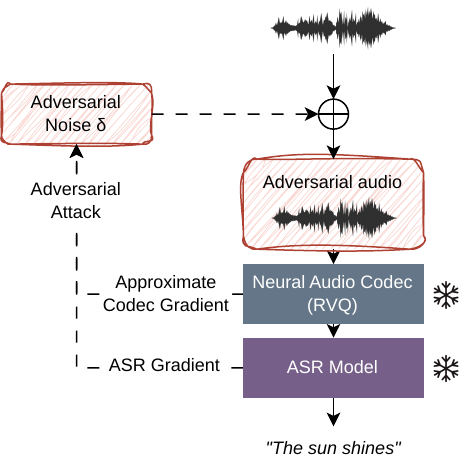}
    \caption{Codec-based inference-time transformation for defended ASR. PGD ignores the codec during optimization, BPDA+EOT adapts with an identity-gradient approximation, and SoftVQ-PGD adapts with a differentiable relaxation of RVQ quantization.}
    \label{fig:codec_pipeline}
\end{figure}

We consider untargeted $\ell_\infty$ bounded waveform attacks that optimize an ASR loss as a surrogate for transcription error under white-box access to the ASR model. The non-adaptive PGD baseline omits the codec during optimization and applies it only during defended evaluation. The adaptive attacks additionally assume knowledge of the full codec-ASR pipeline and differ in how they approximate gradients through the non-differentiable RVQ bottleneck: BPDA+EOT uses an identity-gradient approximation, while SoftVQ-PGD uses a differentiable relaxation of RVQ quantization. The general pipeline is illustrated in Fig.~\ref{fig:codec_pipeline}.


\vspace{1mm} \noindent
\textbf{Projected Gradient Descent (PGD).}
We use PGD~\cite{madry2018towards} as the non-adaptive white-box baseline. Perturbations are iteratively updated using the ASR loss gradient and projected onto the $\ell_\infty$ ball, ignorant of codec defense.

\vspace{1mm} \noindent
\textbf{Backward Pass Differentiable Approximation with Expectation over
Transformation (BPDA+EOT).}
For adaptive evaluation, we first apply BPDA+EOT~\cite{athalye2018obfuscated}. Perturbations are optimized through the codec transformation $C(\cdot)$, with the gradient of $C$ approximated by the identity during backpropagation. EOT performs Monte Carlo averaging of the loss over Gaussian-perturbed inputs prior to quantization, smoothing the objective despite the discontinuities at RVQ decision boundaries. Projected gradient optimization then proceeds under the same $\ell_\infty$ constraint.

\vspace{1mm} \noindent
\textbf{Soft Vector Quantization PGD (SoftVQ-PGD).}
Because BPDA+EOT uses an identity surrogate rather than differentiating through codebook selection, we additionally evaluate SoftVQ-PGD, which instantiates a quantization-aware gradient surrogate using a differentiable relaxation of hard vector quantization~\cite{agustsson2017soft,jang2017categorical,bengio2013estimating}.

This relaxation is used only during attack optimization and does not modify the hard-quantized codec used for final defended evaluation. For each RVQ layer with residual input $r_k$ and codebook
$\mathcal{C}_k=\{e_{k,j}\}_{j=1}^{M_k}$, hard nearest-neighbor
assignment is replaced during attack optimization by a temperature-scaled
soft assignment over codebook distances:
\begin{equation}
w_{k,j}(r_k) =
\frac{\exp\left(-\|r_k-e_{k,j}\|_2^2 / \tau\right)}
{\sum_{\ell=1}^{M_k}\exp\left(-\|r_k-e_{k,\ell}\|_2^2 / \tau\right)} ,
\end{equation}
where $\tau$ controls the sharpness of the relaxation. 

As $\tau \rightarrow 0$, the weights concentrate on the nearest codeword and recover hard assignment, while larger $\tau$ distributes mass across multiple codewords. The relaxed quantized vector is consequently:
\begin{equation}
\tilde{Q}_k(r_k)=\sum_{j=1}^{M_k} w_{k,j}(r_k)e_{k,j}.
\end{equation}

The relaxed quantizers follow the standard RVQ residual sequence, $r_{k+1}=r_k-\tilde{Q}_k(r_k)$, producing a relaxed latent representation that is passed through the codec decoder and ASR model so gradients propagate through codebook-distance geometry rather than an identity approximation. PGD is then performed on the input waveform under the same $\ell_\infty$ constraint as the other attacks, while $\tau$ is gradually decreased over optimization so the soft assignments approach hard nearest-neighbor quantization.

\section{Experimental Setup}

\noindent \textbf{Dataset and ASR models.}
Experiments use random samples from LibriSpeech test-clean~\cite{librispeech}. To increase speaker diversity, we sample utterances approximately evenly across speakers rather than drawing all utterances uniformly. PGD is evaluated on 1000 samples. Because adaptive attacks are more
computationally expensive, BPDA+EOT and SoftVQ-PGD are evaluated on a
500-sample subset of the PGD evaluation set. We evaluate the codec-ASR pipeline on Whisper (base)~\cite{radford2022robustspeechwhisper} and wav2vec~2.0 (base)~\cite{baevski2020wav2vec20}.

\vspace{1mm} \noindent \textbf{Neural codec configurations.}
We evaluate pretrained EnCodec~\cite{defossez2022high}, DAC~\cite{kumar2023high} (24~kHz), and Mimi~\cite{defossez2024moshi} without retraining, covering a range of frame rates, codebook depths, and acoustic-versus-semantic learning objectives. PGD, BPDA+EOT, and SoftVQ-PGD are evaluated at depths $N \in \{2,4,6,8,10,12,16,24,32\}$, providing a common sweep from shallow, strongly compressed settings to deeper, higher-capacity reconstructions. The codecs are pretrained and not finetuned for ASR on this dataset.

\vspace{1mm} \noindent \textbf{Baselines.}
We compare against MP3 and Opus compression, two classical compression-based transformations used in prior adversarial defense studies~\cite{hussain2021waveguard,chen2023towards,raber2025keep}. MP3 and Opus are evaluated at 4.5 kbps, matching the bitrate of EnCodec and DAC (6 codebooks, 4.5 kbps) and the maximum available bitrate of Mimi (32 codebooks, 4.4 kbps). This controls nominal bitrate so that observed differences cannot be attributed to the requested compression rate alone, while allowing codec-specific rate--distortion behavior to differ.

\vspace{1mm} \noindent \textbf{Adversarial attacks.}
We evaluate three untargeted $\ell_\infty$-bounded attack formulations with 100 optimization iterations: non-adaptive PGD~\cite{madry2018towards}, adaptive BPDA+EOT~\cite{athalye2018obfuscated,athalye2018synthesizing}, and adaptive SoftVQ-PGD. PGD and SoftVQ-PGD use $\epsilon \in \{0.0001, 0.001, 0.005, 0.01, 0.02\}$. Due to computational cost, BPDA+EOT is evaluated at $\epsilon \in \{0.005, 0.01, 0.02\}$. 
We focus on untargeted $\ell_\infty$ attacks, leaving targeted attacks and alternative norm constraints for future work.

All three attacks use the same projected sign-PGD waveform update and differ only in how gradients are computed. We use step size $\alpha=\epsilon/40$, initialize $\delta \sim \mathcal{U}(-\epsilon,\epsilon)$, and use no restarts or gradient normalization beyond the sign update. After each step, $\delta$ is projected to $[-\epsilon,\epsilon]$ and the waveform is clipped to $[-1,1]$. Clean audio is peak-normalized to unit amplitude before attack generation. Whisper gradients maximize teacher-forced token cross-entropy over decoder logits, while wav2vec~2.0 gradients maximize CTC loss against the normalized transcript. Non-adaptive PGD computes these gradients through the ASR model alone, applying the selected codec only during defended evaluation.

Following~\cite{athalye2018obfuscated}, BPDA+EOT uses codec reconstruction in the forward pass and approximates codec gradients with a straight-through identity function. EOT averages the loss over $K=8$ Gaussian-jittered codec forwards with $\sigma=0.001$. Preliminary 50-sample sweeps over EOT samples, jitter scale, and iteration count did not change the qualitative trends, so we use this configuration in the main BPDA+EOT experiments.


For SoftVQ-PGD, we use the same projected sign-PGD update as PGD, replacing hard RVQ assignment with a differentiable soft assignment during optimization, following soft-to-hard vector quantization relaxations~\cite{agustsson2017soft,bengio2013estimating}. We do not use EOT. The temperature $\tau$ is linearly annealed from approximately $1.0$ to $0.1$ for all codecs and depths, sharpening assignments toward hard quantization. Final evaluation always uses the original hard-quantized codec.

SoftVQ-PGD depends on this schedule, whose effective sharpness can vary across codecs, depths, and latent scales. We therefore treat it as a fixed quantization-aware evaluation rather than tune it per configuration. The results are not worst-case estimates over relaxation schedules. Instead, they show that an RVQ-aware surrogate is sufficient to substantially alter the apparent RVQ-depth robustness trend relative to identity-gradient BPDA.

\vspace{1mm} \noindent \textbf{Uniform noise control.}
We evaluate uniform noise over the full $\epsilon$ grid as a non-adversarial calibration control, sampling $\delta \sim \mathcal{U}(-\epsilon,\epsilon)$ without gradient optimization and applying the same clipping as the attacks. This separates bounded signal distortion from adversarial degradation.

\vspace{1mm} \noindent \textbf{Evaluation metrics.}
We report word error rate (WER) for Whisper and wav2vec~2.0 after defense reconstruction. Reported confidence intervals are 95\% utterance-level bootstrap intervals. PESQ and STOI are auxiliary distortion measures, computed between the defended waveform and the clean input.




\section{Results and Discussion}

We organize the results as a progression from calibration to increasingly adaptive evaluation. First, we verify that the perturbation budgets induce adversarial rather than merely noisy degradation by comparing PGD with uniform noise in the absence of codec reconstruction. We then use non-adaptive PGD to characterize how RVQ depth affects defended ASR when the codec is applied only at inference, before evaluating BPDA+EOT as a standard adaptive baseline and SoftVQ-PGD as a quantization-aware adaptive attack. This sequence asks whether the apparent RVQ-depth optimum is stable across threat models, or instead depends on the attack surrogate used to optimize the defended codec-ASR pipeline.


\subsection{Attack Calibration and Non-Adaptive Depth Trade-off}
\label{sec:adv_across_depth}

\begin{table}[!t]
\centering
\caption{No-codec perceptual quality and mean WER (\%) on
LibriSpeech test-clean ($n=1{,}000$). PESQ/STOI are computed relative to
clean audio. For PGD rows, perceptual metrics are computed on the ASR-specific adversarial waveforms.}
\label{tab:attack_sanity}
\scriptsize
\setlength{\tabcolsep}{4pt}
\renewcommand{\arraystretch}{0.96}
\begin{tabular}{@{}llcc cc@{}}
\toprule
& & \multicolumn{2}{c}{Whisper} & \multicolumn{2}{c}{wav2vec~2.0} \\
\cmidrule(lr){3-4} \cmidrule(l){5-6}
$\epsilon$ & Pert. & PESQ/STOI & WER & PESQ/STOI & WER \\
\midrule
0.0    & Clean   & 4.64 / 1.00   & 5.25  & 4.64 / 1.00   & 3.94 \\
\midrule
0.0001 & Noise & 4.64 / 1.00   & 5.28  & 4.64 / 1.00   & 3.89 \\
       & PGD   & 4.63 / 1.00   & 6.75  & 4.63 / 1.00   & 5.77 \\
\midrule
0.001  & Noise & 4.30 / 0.999  & 5.21  & 4.30 / 0.999  & 3.89 \\
       & PGD   & 4.12 / 0.997  & 38.46 & 4.24 / 0.998  & 10.06 \\
\midrule
0.005  & Noise & 2.97 / 0.993  & 5.51  & 2.97 / 0.993  & 4.02 \\
       & PGD   & 2.85 / 0.982  & 76.78 & 2.90 / 0.988  & 17.38 \\
\midrule
0.01   & Noise & 2.23 / 0.983  & 5.60  & 2.23 / 0.983  & 4.43 \\
       & PGD   & 2.16 / 0.965  & 82.09 & 2.19 / 0.975  & 23.23 \\
\midrule
0.02   & Noise & 1.63 / 0.962  & 6.14  & 1.63 / 0.962  & 5.23 \\
       & PGD   & 1.59 / 0.932  & 95.52 & 1.60 / 0.947  & 34.12 \\
\bottomrule
\end{tabular}
\end{table}

We first calibrate the perturbation budgets in the absence of a codec defense. Table~\ref{tab:attack_sanity} shows that uniform noise at the same $\ell_\infty$ budgets leaves WER near the clean baseline, while PGD produces large WER increases for both ASR models. Because uniform noise and PGD have comparable PESQ/STOI at each budget, the undefended degradation is attributable to adversarial structure rather than bounded signal distortion alone. This establishes PGD as a strong base-ASR attack before evaluating codec transformations.

\begin{figure}[!t]
    \centering
    \includegraphics[width=\linewidth]{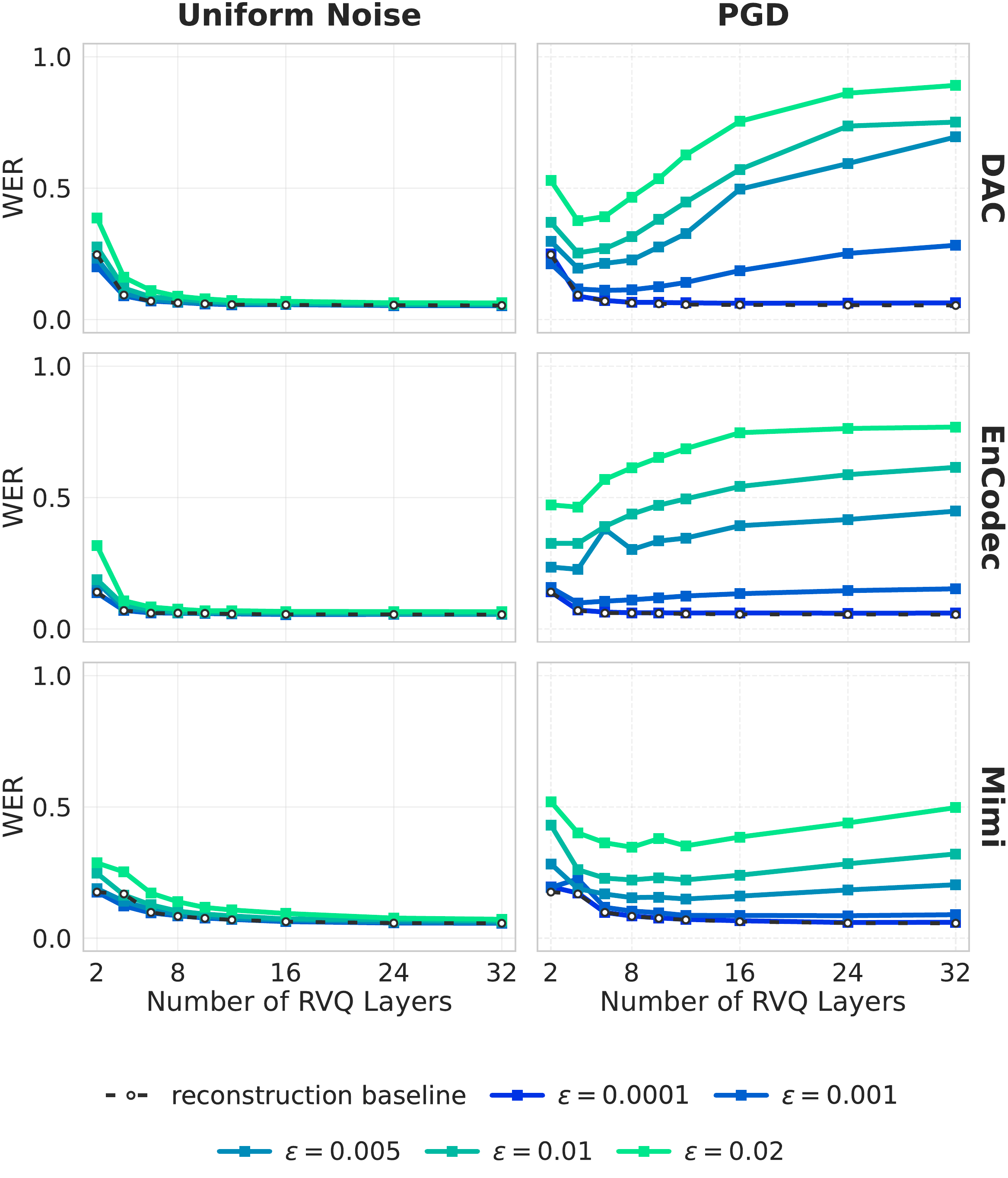}
    \caption{Defended WER versus RVQ depth $N$ under uniform noise and PGD for DAC, EnCodec, and Mimi, evaluated with Whisper. Uniform noise tracks the clean codec baseline across depths, while PGD causes larger, depth-dependent degradation.}
    \label{fig:capacity_depth}
    \vspace{-3mm}
\end{figure}


\begin{table}[!t]
\centering
\caption{WER (\%, mean $\pm$ 95\% CI) under PGD ($\epsilon=0.001$), with PESQ/STOI means. All codecs are evaluated at 4.5 kbps, except Mimi at 4.4 kbps and MP3 at 8 kbps. For neural codecs, cb = codebooks.}
\label{tab:best_codecs}
\scriptsize
\setlength{\tabcolsep}{2.5pt}
\renewcommand{\arraystretch}{1.1}
\begin{tabular}{l cc cc}
\toprule
& \multicolumn{2}{c}{Whisper} & \multicolumn{2}{c}{wav2vec~2.0} \\
\cmidrule(lr){2-3} \cmidrule(lr){4-5}
Defense & WER (\%) $\downarrow$ & PESQ/STOI & WER (\%) $\downarrow$ & PESQ/STOI \\
\midrule
None          & 38.46 $\pm$ 5.15 & 4.12/0.997 & 10.06 $\pm$ 0.73 & 4.24/0.998 \\
MP3           & 19.59 $\pm$ 1.28 & 1.50/0.843 & 17.51 $\pm$ 1.08 & 1.50/0.844 \\
Opus          & 19.22 $\pm$ 1.25 & 1.39/0.726 & 16.66 $\pm$ 1.04 & 1.39/0.728 \\
\midrule
EnCodec (6cb) & 10.53 $\pm$ 0.84 & 2.47/0.924 & 5.67 $\pm$ 0.53 & 2.51/0.925 \\
DAC (6cb)     & 11.07 $\pm$ 0.86 & 2.93/0.937 & 6.28 $\pm$ 0.57 & 2.94/0.938 \\
Mimi (32cb)   & \textbf{8.94 $\pm$ 0.74} & 3.33/0.960 & \textbf{5.17 $\pm$ 0.49} & 3.35/0.960 \\
\bottomrule
\end{tabular}
\vspace{-3mm}
\end{table}

We next evaluate RVQ depth under non-adaptive PGD. In this setting, the attack is optimized against the ASR model alone, and the codec is applied only during defended evaluation. Fig.~\ref{fig:capacity_depth} shows that the same calibration pattern holds under codec reconstruction: uniform noise tracks the clean baseline across RVQ depths, while PGD produces substantially larger, depth-dependent WER increases. Under PGD, shallow depths can over-compress speech, intermediate depths often reduce transcription error, and larger depths increasingly preserve attack-relevant structure. This produces an apparent robustness trade-off under simple non-adaptive evaluation.

Table~\ref{tab:best_codecs} compares representative codec settings against bitrate-matched Opus and minimum-bitrate MP3 baselines at $\epsilon=0.001$. Neural codecs achieve the lowest WER for both ASR models. At matched bitrate, neural codecs outperform Opus in both WER and reconstruction quality, while also beating MP3 on both measures despite its higher 8-kbps minimum bitrate. Thus, bitrate alone cannot explain their gains. Together with the depth sweep in Fig.~\ref{fig:capacity_depth}, these results show that RVQ-based codec transformations can provide practical robustness gains under non-adaptive PGD, but this apparent depth optimum depends on the attack setting, as shown by the following adaptive evaluations.

\subsection{BPDA+EOT Optimization Through RVQ}
\label{sec:bpda_results}

\begin{table}[!t]
\centering
\caption{Mean WER (\%) $\pm$ 95\% CI under BPDA+EOT. All codecs are evaluated at 4.5 kbps, except Mimi at 4.4 kbps and MP3 at 8 kbps. For neural codecs, cb = codebooks}
\label{tab:best_codecs_bpda_eot8}
\scriptsize
\setlength{\tabcolsep}{5pt}
\begin{tabular}{llcc}
\toprule
$\epsilon$ & Method & Whisper & wav2vec 2.0 \\
\midrule
\multirow{5}{*}{0.005}
  & MP3                       & 59.15 $\pm$ 2.32 & 21.50 $\pm$ 1.59 \\
  & Opus                      & 21.87 $\pm$ 1.89 & 21.85 $\pm$ 1.69 \\
\cmidrule(l){2-4}
  & EnCodec (6cb)             & 7.10 $\pm$ 1.00 & 6.93 $\pm$ 0.92 \\
  & DAC (6cb)                 & \textbf{7.03 $\pm$ 1.12} & 5.66 $\pm$ 0.81 \\
  & Mimi (32cb)               & 7.04 $\pm$ 1.03 & \textbf{5.22 $\pm$ 0.71} \\
\midrule
\multirow{5}{*}{0.01}
  & MP3                       & 104.69 $\pm$ 17.20 & 24.15 $\pm$ 1.68 \\
  & Opus                      & 32.57 $\pm$ 2.11 & 29.94 $\pm$ 1.94 \\
\cmidrule(l){2-4}
  & EnCodec (6cb)             & 11.15 $\pm$ 1.33 & 10.01 $\pm$ 1.10 \\
  & DAC (6cb)                 & \textbf{8.70 $\pm$ 1.16} & \textbf{7.76 $\pm$ 0.98} \\
  & Mimi (32cb)               & 10.77 $\pm$ 1.30 & 7.81 $\pm$ 0.91 \\
\midrule
\multirow{5}{*}{0.02}
  & MP3                       & 107.46 $\pm$ 14.08 & 30.14 $\pm$ 1.95 \\
  & Opus                      & 55.08 $\pm$ 9.59 & 46.97 $\pm$ 2.19 \\
\cmidrule(l){2-4}
  & EnCodec (6cb)             & 24.66 $\pm$ 1.88 & 18.84 $\pm$ 1.53 \\
  & DAC (6cb)                 & \textbf{16.09 $\pm$ 1.79} & 14.23 $\pm$ 1.45 \\
  & Mimi (32cb)               & 23.15 $\pm$ 1.81 & \textbf{13.52 $\pm$ 1.21} \\
\bottomrule
\end{tabular}
\vspace{-3mm}
\end{table}

\begin{figure}[!t]
    \centering
    \includegraphics[width=\linewidth]{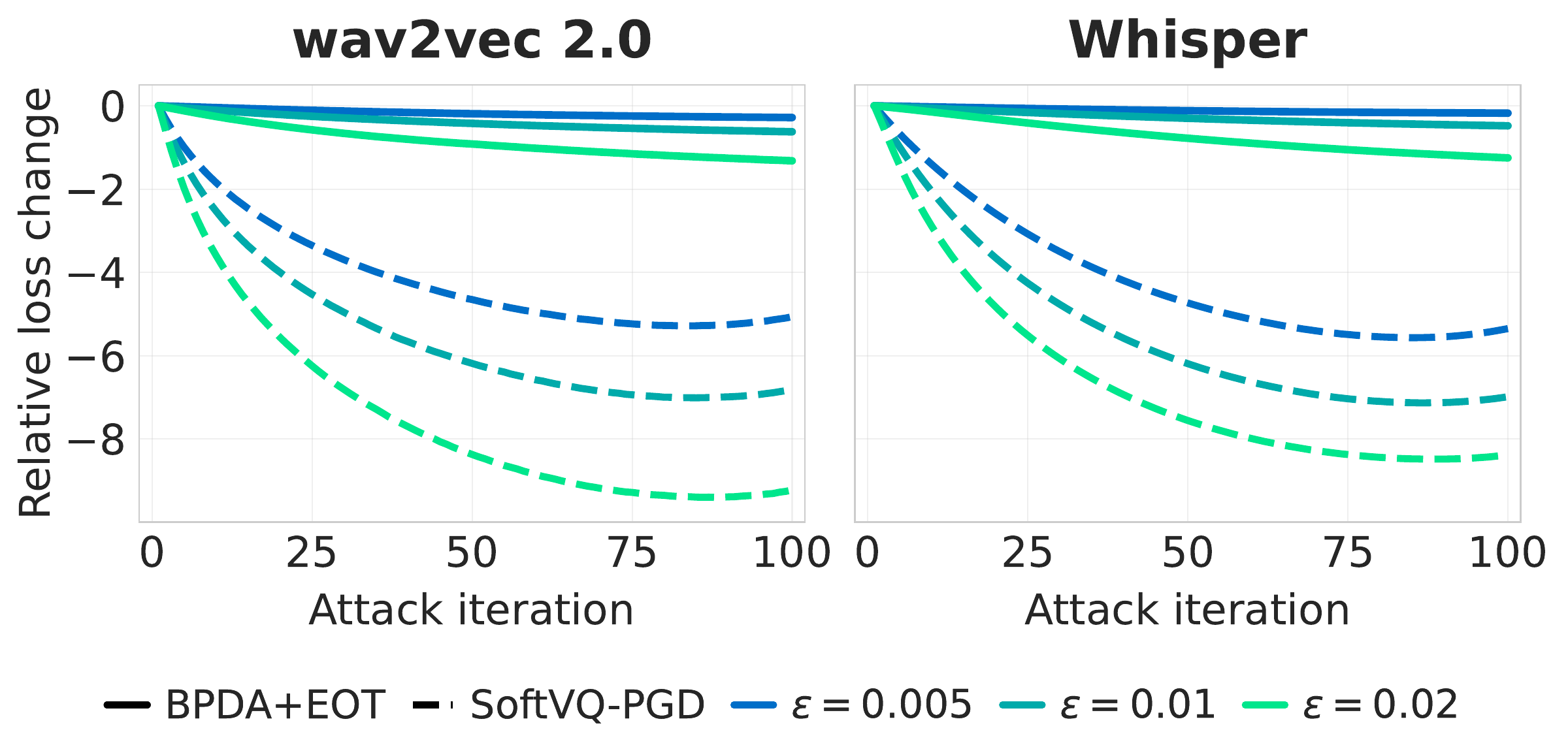}
    \caption{Relative optimization-objective change for BPDA+EOT and SoftVQ-PGD, averaged across codec families and RVQ depths, normalized within ASR models. SoftVQ-PGD shows greater optimization progress.}
    \label{fig:gradient_masking}
    \vspace{-3mm}
\end{figure}

We next evaluate BPDA+EOT as an adaptive baseline for the defended codec-ASR pipeline. Table~\ref{tab:best_codecs_bpda_eot8} reports BPDA+EOT WER. Under this evaluation, RVQ-based codec transformations appear substantially more effective than both MP3 and Opus. Across all budgets, neural codecs yield lower WER than classical compression baselines. However, the optimization diagnostics below suggest that these WER values should be interpreted as BPDA+EOT outcomes rather than final adaptive robustness estimates.

To diagnose this, Fig.~\ref{fig:gradient_masking} compares two adaptive attacks through the defended codec-ASR pipeline: BPDA+EOT, which uses an identity-gradient approximation, and SoftVQ-PGD, which uses a quantization-aware RVQ relaxation. BPDA+EOT shows only modest objective movement even at larger perturbation budgets, whereas SoftVQ-PGD makes substantially greater optimization progress for both ASR models. This suggests that BPDA+EOT can under-optimize through the RVQ bottleneck, so we treat it as a useful adaptive baseline rather than a definitive robustness estimate, motivating the SoftVQ-PGD evaluation in Section~\ref{sec:softvq_results}.

\subsection{SoftVQ-PGD Depth Trends}
\label{sec:softvq_results}
\vspace{-1.5mm}

\begin{figure}[!t]
    \centering

    \includegraphics[width=\linewidth]{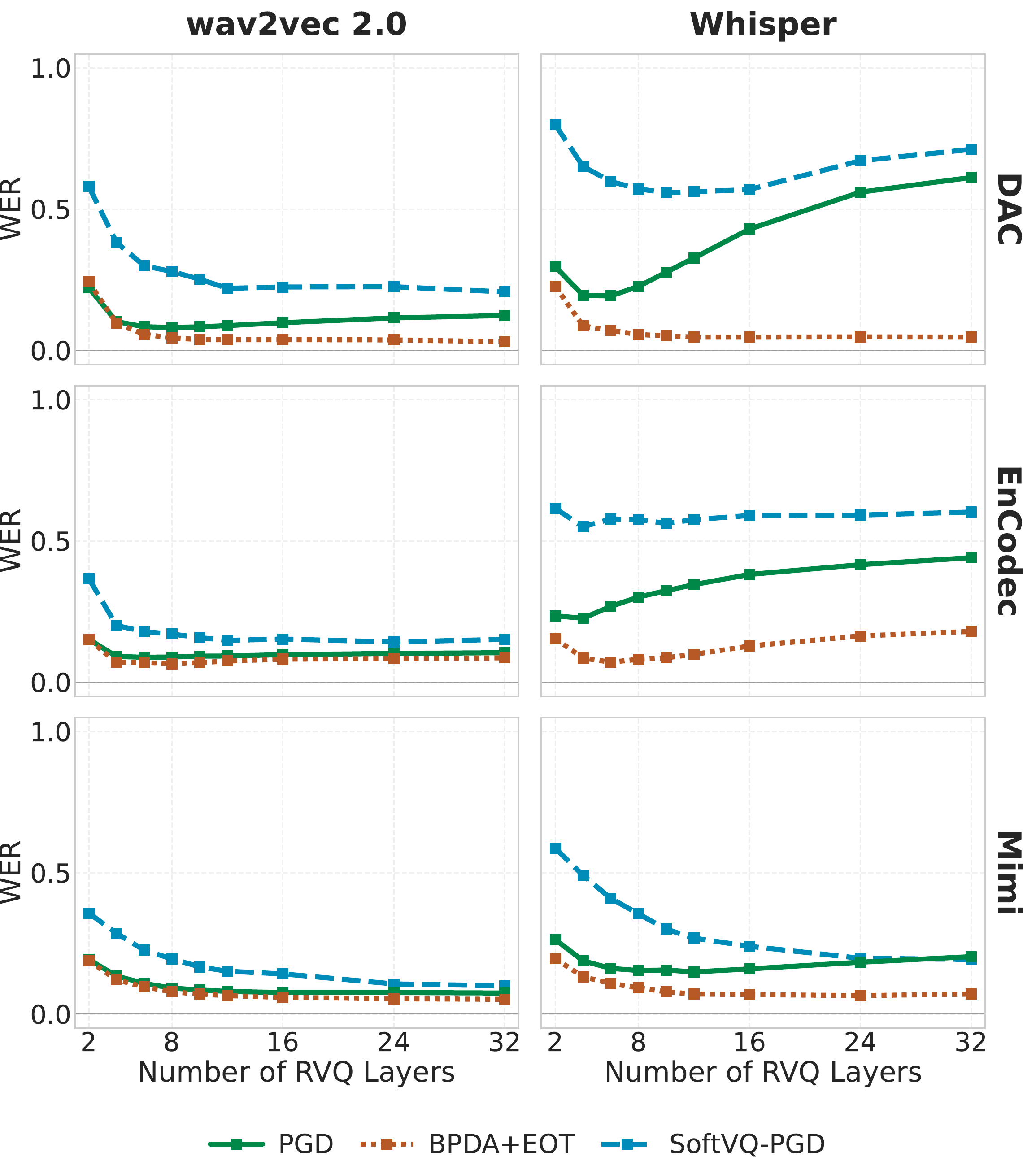}
    \caption{Defended WER versus RVQ depth $N$ at $\epsilon=0.005$ under PGD, BPDA+EOT, and SoftVQ-PGD across ASR models and codec families. SoftVQ-PGD generally yields the highest defended WER, indicating stronger optimization through the RVQ-based defense than PGD or BPDA+EOT.}
    \label{fig:attack_wer_depth_comparison}
\end{figure}

\begin{figure}[!t]
    \centering
    \includegraphics[width=\linewidth]{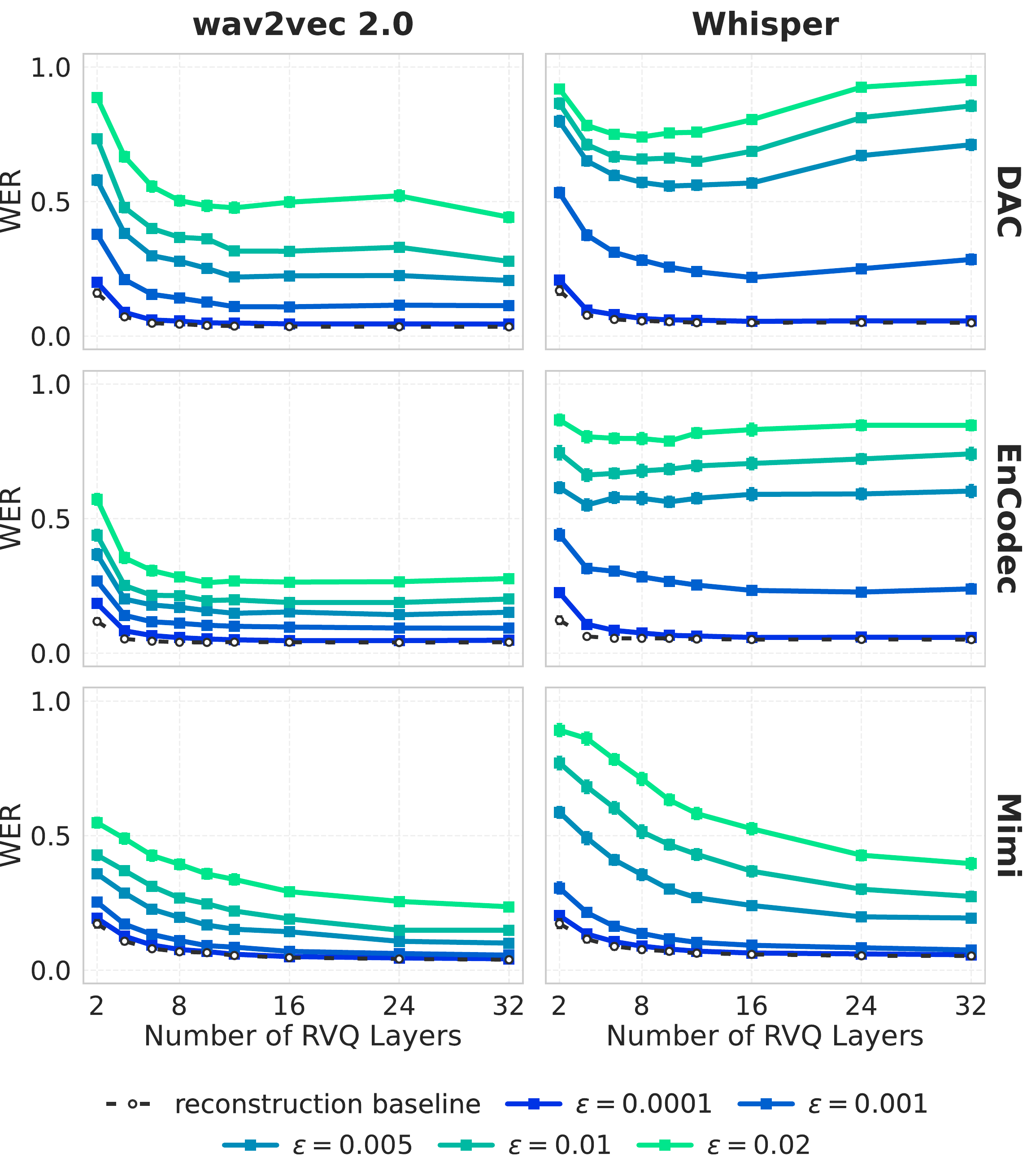}
    \caption{Defended WER under SoftVQ-PGD versus RVQ depth for wav2vec~2.0 and Whisper across DAC, EnCodec, and Mimi. SoftVQ-PGD depth trends flatten, decrease, or vary by codec-ASR pair.}
    \label{fig:softvq_wer_depth}
    \vspace{-4mm}
\end{figure}

Having diagnosed weak optimization under BPDA+EOT, we evaluate SoftVQ-PGD as a quantization-aware adaptive attack. Fig.~\ref{fig:attack_wer_depth_comparison} compares defended WER across attacks at $\epsilon=0.005$. Relative to BPDA+EOT and non-adaptive PGD, SoftVQ-PGD produces substantially higher WER at shallow and intermediate depths, changing the depth-dependent pattern observed under the other attack formulations. This behavior is consistent with the different optimization problem solved by SoftVQ-PGD. Non-adaptive PGD optimizes only against the ASR model, so the codec can distort part of the adversarial structure. SoftVQ-PGD instead optimizes through a relaxed approximation of the codec, allowing optimization to produce perturbations that remain harmful after RVQ reconstruction.

Sweeping SoftVQ-PGD across perturbation budgets shows that the non-adaptive PGD depth trend from Section~\ref{sec:adv_across_depth} does not transfer cleanly to the quantization-aware attack. Whereas several PGD settings exhibit an intermediate-depth minimum followed by increasing WER at larger depths, Fig.~\ref{fig:softvq_wer_depth} shows that SoftVQ-PGD often decreases or flattens with additional RVQ layers, with exceptions such as DAC with Whisper. Thus, the intermediate depth optimum observed under non-adaptive PGD is not an intrinsic property of RVQ. Instead, it reflects an attack-dependent WER-depth trade-off that changes when the attacker optimizes through a quantization-aware surrogate.

Overall, these results show that codec robustness conclusions depend on the attack formulation. Non-adaptive PGD suggests that RVQ depth can create a useful compression--fidelity trade-off, and BPDA+EOT appears to preserve much of this robustness. However, the optimization diagnostics and SoftVQ-PGD results show that this conclusion is not stable under quantization-aware adaptive evaluation. 
\vspace{-0.5mm}

\section{Conclusion}
\vspace{-1.5mm}

We analyzed RVQ-based neural audio codecs as inference-time transformations for untargeted $\ell_\infty$ attacks on ASR. 
RVQ showed a U-shaped robustness depth trend under non-adaptive PGD, while BPDA+EOT suggested even greater robustness due to poor optimization. In contrast, SoftVQ-PGD caused much more degradation and largely removed the intermediate-depth optimum.
These results show that robustness trends across RVQ-depth vary by attack, not just codec architecture.
These conclusions are limited to untargeted $\ell_\infty$ attacks on LibriSpeech test-clean with base Whisper and wav2vec~2.0. Future work will explore targeted and alternative-norm attacks, broader ASR models, and additional datasets.

\vspace{-0.5mm}

\section*{Acknowledgment}
\vspace{-1.5mm}

This work was supported by the National Science Foundation Graduate Research Fellowship (NSF GRFP) grant DGE-1842487, as well as the Office of the Director of National Intelligence (ODNI), Intelligence Advanced Research Projects Activity (IARPA), via the ARTS Program under contract D2023-2308110001. This work used NCSA Delta GPUs through allocation CIS261094 from the Advanced Cyberinfrastructure Coordination Ecosystem: Services \& Support (ACCESS) program~\cite{nsf}, which is supported by U.S. NSF grants \#2138259, \#2138286, \#2138307, \#2137603, and \#2138296.

\section*{AI-Generated Content Disclosure}
Generative AI tools were used to assist with table and figure formatting and language refinement of this manuscript. All experimental design, implementation, data collection, analysis, and manuscript preparation were conducted by the authors.

\bibliographystyle{IEEEtran}
\bibliography{mybib}

\end{document}